\documentclass[aps,pra, onecolumn, superscriptaddress]{revtex4}

\usepackage{graphicx}
\usepackage{bm}
\usepackage{amsmath}
\usepackage[colorlinks=true,allcolors=blue]{hyperref}
\usepackage{amssymb}
\usepackage[labelformat=simple]{subfig}

\usepackage{lipsum}

\providecommand{\be}{\begin{equation}}
\providecommand{\ee}{\end{equation}}
\providecommand{\ba}{\begin{eqnarray}}
\providecommand{\ea}{\end{eqnarray}}

\usepackage{mathbbol}
\usepackage[makeroom]{cancel}

\begin{document}

\title{Time-Dependent Dephasing and Quantum Transport}

\author{Saulo V. Moreira}
\affiliation{Department of Physics and NanoLund, Lund University, Box 118, 22100 Lund, Sweden}
\affiliation{Centro de Ci\^encias Naturais e Humanas, Universidade Federal do ABC - UFABC, Santo Andr\'e, Brazil}
\author{Breno Marques}
\affiliation{Centro de Ci\^encias Naturais e Humanas, Universidade Federal do ABC - UFABC, Santo Andr\'e, Brazil}
\author{Fernando L. Semi\~ao}
\affiliation{Centro de Ci\^encias Naturais e Humanas, Universidade Federal do ABC - UFABC, Santo Andr\'e, Brazil}

\begin{abstract}

The investigation of the phenomenon of dephasing assisted quantum transport, which happens when the presence of dephasing benefits the efficiency of this process, has been mainly focused on Markovian scenarios associated with constant and positive dephasing rates in their respective Lindblad master equations.
What happens if we consider a more general framework, where time-dependent dephasing rates are allowed, thereby permitting the possibility of non-Markovian scenarios? Does dephasing assisted transport still manifest for non-Markovian dephasing?
Here, we address these open questions in a setup of coupled two-level systems.
Our results show that the manifestation of non-Markovian dephasing assisted transport depends on the way in which the incoherent energy sources are locally coupled to the chain.
This is illustrated with two different configurations, namely non-symmetric and symmetric.
Specifically, we verify that non-Markovian dephasing assisted transport manifested only in the non-symmetric configuration.
This allows us to draw a parallel with the conditions in which time-independent Markovian dephasing assisted transport manifests.
Finally, we find similar results by considering a controllable and experimentally implementable system, which highlights the significance of our findings for quantum technologies.
 
\end{abstract}
\pacs{}

\maketitle

\section{Introduction}

Dephasing-assisted transport means currents enhanced by dephasing~\cite{PlenioTransport, Guzik}.
This implies that  open system dynamics may surpass the correspondent unitary evolution in terms of transport efficiency.
On the one hand, this defied the notion that, in~general, the~presence of noise tends to jeopardize the efficiency of tasks performed by quantum systems~\cite{super}.
 
On the other hand, it helped us to understand energy transport behavior in quantum systems subject to heavily noisy conditions in harsh natural environments, which  shows an outstanding ability to effectively transfer energy.
A paradigmatic example  is the widely studied Fenna--Mathew--Olson (FMO) complex, a~structure present in green sulphur bacteria that channels the energy captured from solar light to a reaction centre~\cite{Engel, Caruso, Nonlin, fmo1, Jang, fmo2}.
As well, the~comprehension of dephasing-assisted transport  is of central importance for quantum technologies.

Indeed, the~possibility of exploiting this to achieve improved transport efficiencies is very appealing from the point of view of practical implementations, principally for quantum technology applications including controlled quantum systems, such as disordered organic semiconductors~\cite{Schachenmayer}, networks of fiber-optic resonators~\cite{Viciani}, reconfigurable networks for the simulation of single-particle quantum transport~\cite{Quiroz},  nuclear magnetic resonance systems~\cite{exp}, and~quantum emitters near a graphene sheet under the influence of a magnetic field~\cite{Abrantes}.

The theoretical studies of dephasing-assisted transport have been mainly focused on time-independent interaction between the system and environment  \citep{Elinor, Semiao, PlenioTransport}.
Therefore, investigations of time-dependent dephasing in a transport scenario that includes more general Markovian as well as non-Markovian evolutions have the potential to drive new applications in the context of quantum technologies~\cite{NMTransport, Ancheyta}.
Furthermore, in~recent years, there has been a great interest in the fundamental and practical aspects of non-Markovianity~\cite{Souza, paulo, fred, mauro, exp, fermi1, fermi2, fermi3, Maier}. With~the tools resulting from these studies and the experimental advances that have been reported, it is natural to envisage new possibilities to exploit such systems in the context of quantum transport.

In this work, we will study how the presence of time-dependent dephasing in a chain of coupled two-level systems affects quantum transport efficiency in Markovian and non-Markovian scenarios. In~doing so, we tackle a relevant question in the field of open quantum systems, which is the impact of time-dependent scenarios on quantum transport. 
We will be using the fact that important examples of non-Markovian evolutions can be characterized by Lindblad-like master equations for which the time dependent decoherence rate achieves negative values~\mbox{\cite{Hall2014, RevMarkov, RHP, BLP, NMTransport}}.

This paper is organized as follows.
First, we review the canonical form of the Lindblad-like master equations and the characterization of non-Markovianity via master equations in Section~\ref{scnm}.
Then, we describe the transport model in Section~\ref{model}.
In Section~\ref{DatTime}, we present our results and analyse the transport efficiency in some time-dependent dephasing scenarios, and~extend the analysis for results obtained in the context of a controlled quantum system in Section~\ref{example}. In~Section~\ref{conc}, we present our~conclusions.

\section{Characterizing Time-Dependent Non-Markovian~Evolutions}\label{scnm}



Time-local master equations~\cite{Lindblad, Breuer1}  can  be expressed in a Lindblad-like form as
\begin{equation}\label{Canonical}
\dot{\rho}(t) = -i[H(t), \rho]   +   \sum_k^{d^2-1} \gamma_k(t) \left(\hat{L}_k(t) \rho \hat{L}_k^\dagger(t)
  - \frac{1}{2}\{ \hat{L}_k^\dagger(t) \hat{L}_k(t), \rho \} \right),
\end{equation}
with a unique  set of functions $\gamma_k(t)$, not necessarily positive for all times~\cite{Hall2014}. Here, $d$ is the dimension of the state space, $H(t)$ is a Hermitian operator, and~$\hat{L}_k(t)$ constitutes an orthonormal basis of traceless operators, i.e.,
\begin{equation}\label{TracelessOperators}
{\rm Tr}[\hat{L}_k(t)] = 0, \ \ {\rm Tr}[\hat{L}_j^\dagger(t)\hat{L}_k(t)] = \delta_{jk}.
\end{equation}

Since any time-local master equation can be written in this canonical form, in~which each $\gamma_k(t)$ is uniquely determined, it turns out that Equation~(\ref{Canonical}) may be used to characterize non-Markovianity~\cite{Hall2014}.
In fact,  $\gamma_k(t) \geq 0$ means Markovianity, since it  is equivalent to the divisibility of the map into completely positive evolutions~\cite{Wolf, BLP, RHP, RevMarkov}.
Therefore, a~strictly negative value of $\gamma_k(t)$, for~some $k$ and at any instant of time $t$, indicates  non-Markovianity. 
The fact that each $\gamma_k(t)$ is unique in Equation~(\ref{Canonical}) motivated the use of
\begin{equation}\label{Quantifier}
f_k(t) \equiv \max[0, -\gamma_k(t)] \geq 0,
\end{equation}
as an indicator of non-Markoviany in the channel $k$ and~its integration in time
\begin{equation}\label{QuantifierInt}
F_k(t, t^\prime) = \int_t^{t^\prime} ds f_k(s),
\end{equation}
as a quantifier of the total amount of non-Markovianity of a given channel $k$ in an interval of time from $t$ to $t^\prime$ \cite{Hall2014}.


In general, $\gamma_k(t)$ must satisfy certain constraints for a completely positive evolution.
For instance, consider a master equation for a two-level system given by
\begin{equation}\label{GeneralME}
\dot{\rho}(t) = -i[H(t), \rho(t)] + \frac{1}{2} \sum_k \gamma_k (t) (\sigma_k \rho(t) \sigma_k - \rho(t)),
\end{equation}
where $\sigma_i$ are Pauli matrices ($\sigma_1 = \sigma^x, \sigma_2 = \sigma^y, \sigma_3  = \sigma^z$), and~$H(t)$ is Hermitian.
Complete positivity of the map, in~the interval from $0$ to $t$, is guaranteed if the following set of conditions are fulfilled~\cite{Hall2008}
\begin{equation}
\Gamma_j + \Gamma_k \leq 1+\Gamma_l,
\end{equation}
for all permutations $j, k , l$ of $1, 2, 3$ where $\Gamma_j \equiv \exp(- \int_0^t ds [\gamma_k(s) + \gamma_l(s)])$.

Let us illustrate this with a simple case where $\gamma_1(t) = \gamma_2(t) = 0$ and $\gamma_3(t) = \gamma(t)$, i.e.,
\begin{equation}\label{GeneralME1}
\dot{\rho}(t) = -i[H(t), \rho(t)] + \frac{1}{2} \gamma(t) (\sigma^z \rho(t) \sigma^z - \rho(t)).
\end{equation}

It is straightforward to show that $\int_0^t \gamma(s)ds \geq 0$ is the requirement for the map to be completely positive.
This master equation will be important for our investigation of the phenomenon of non-Markovian dephasing-assisted transport in the rest of the paper.
An example is given by $\gamma(t) =  \gamma\sin(\nu t)$, where $\nu$ is integer and $\gamma \geq 0$ \cite{Darius}.
Such functions satisfy the aforementioned condition and, therefore, the map is CP for all $t$.


\section{The~Model}\label{model}

We consider a linear chain of $N$ two-level systems in a first-neighbour coupling model, whose Hamiltonian is given by ($\hbar=1$)
\begin{equation}\label{H}
H = \sum_{i=1}^{N} \frac{\omega_i}{2}\sigma_i^z + \sum_{i=1}^{N-1} \lambda_i(\sigma_i^+\sigma_{i+1}^- + \sigma_{i+1}^+\sigma_i^- ),
\end{equation}
where $\sigma_i^+$ is the operator causing transition from ground to excited state in site $i$, \mbox{$\sigma_i^-=(\sigma_i^+)^\dag$}, $\sigma^z_i$ and $\omega_i$ are the Pauli $z$ operator and the energy associated with $i$th site, respectively, and~$\lambda_i$ is the coupling constant between sites $i$ and $i+1$. 
This model has been extensively used to describe quantum transport, and~this kind of interaction can be implemented, for~instance, in~the context of trapped ions and circuit QED~\cite{Maier, Hauke, Casparis}.


In turn, the~chain is considered to be locally coupled to incoherent energy sources, responsible for incoherent injection and extraction of energy.
More specifically, we consider energy injection at site $1$ and extraction at site $k$, where $2\leq k \leq N$. 
This situation is described by the following terms, to~be added to the master equation
\begin{eqnarray}\label{InOut}
\mathcal{L}_{\text{inj}}\rho &=&  \frac{1}{2} \kappa_{\text{inj}}(2\sigma_1^+ \rho \sigma_1^- - \sigma_1^-\sigma_1^+ \rho - \rho \sigma_1^-\sigma_1^+),\nonumber\\
\mathcal{L}_{\text{ext}}\rho &=&  \frac{1}{2} \kappa_{\text{ext}}(2\sigma_k^- \rho \sigma_k^+ - \sigma_k^+\sigma_k^- \rho - \rho \sigma_k^+\sigma_k^-),
\end{eqnarray}
where $\kappa_{\text{inj}}\,(\kappa_{\text{ext}})$ describes the rate of injection (extraction) of energy into (out of) the chain.
In order to simplify the notation, we omitted the time-dependence of $\rho(t)$ in Equation \eqref{InOut}. 
From now on, we will adopt this simplified~notation.

Finally, we consider that each site is also subjected to local dephasing.
This assumption of local coupling to the environment is reasonable for a weak intercoupling strength between the sites of the chain when compared to the local frequencies~\cite{Gonzalez, Hofer, DeChiara, Mitchison, McConnella, Santos}.
For the sake of simplicity, we will assume that each site is subjected to equivalent dephasing environments.
Therefore, the~total dephasing to which the chain is subjected to is given by
\begin{equation}\label{GeneralME1}
\mathcal{L}_{\text{deph}}\rho = \sum_i^N  \frac{1}{2} \gamma(t) (\sigma_i^z \rho \sigma_i^z - \rho).
\end{equation}

Then, non-Markovianity is the result of $\gamma(t)$ assuming negative values.
 Finally, the~total master equation representing the evolution of the system will read
\begin{equation}\label{Master}
\dot{ \rho} =-i [H,\rho] + \mathcal{L}_{\text{deph}}\rho+\mathcal{L}_{\text{inj}}\rho + \mathcal{L}_{\text{ext}}\rho.
\end{equation}


The reason why we do not consider a general local environment as in Equation~(\ref{GeneralME}), which contains all Pauli matrices, is that we want to have a fair comparison among cases where the only way energy can enter or leave the chain is by means of the same mechanism of incoherent injection and extraction of energy caused by Equation~(\ref{InOut}).
\cite{Liu} has shown that it is possible to simulate arbitrary pure-dephasing dynamics of a qubit through a generic dephasing simulator for one-qubit dephasing. In~this way, it is important to point out that all time-dependent dephasing models discussed here would be experimentally implementable using the results tools of \cite{Liu}.

For the investigation of transport efficiency,  we will consider the stationary value of the rate of variation of the total number operator $\hat{N}=\sum_i\sigma_i^+\sigma_i^-$. In~a broad sense, it can be called the exciton current. In~the stationary state, one finds \mbox{$ {\rm Tr}[\hat{N}\dot\rho]={\rm Tr}[\hat{N}\mathcal{L}_{\text{in}}\rho]+{\rm Tr}[\hat{N}\mathcal{L}_{\text{ext}}\rho]=0$}. Consequently, ${\rm Tr}[\hat{N}\mathcal{L}_{\text{in}}\rho]=-{\rm Tr}[\hat{N}\mathcal{L}_{\text{ext}}\rho]$. As~a figure of merit for the transport efficiency, we then consider $J_{\hat{N}}=|{\rm Tr}[\hat{N}\mathcal{L}_{\text{ext}}\rho]|= \kappa_{\text ext} \,p_{\text{ext}}(\infty)$, where $p_{\text{ext}}(\infty)$ is the asymptotic population of the extraction site. To~be more specific, we will be evaluating the rescaled current \linebreak $\tilde{J}_{\hat{N}} \equiv (\kappa_\text{ext} N)^{-1}  J_{\hat{N}}$. The~basic transport models used here are detailed in the review~\cite{SM}.

\section{Time-Dependent Dephasing Assisted~Transport}\label{DatTime}



\begin{figure}[h!]
 \centering
\includegraphics[scale=0.5]{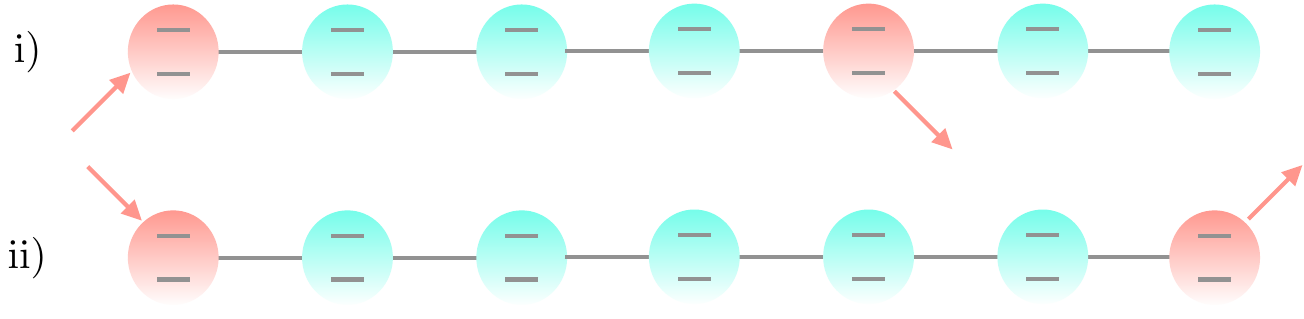}
\label{fig:fig14}
\caption{Representation of the networks considered in this work (\textbf{i}) non-symmetric and (\textbf{ii}) symmetric configurations. The~injection and extraction sites for each configuration are indicated by the ingoing and outgoing arrows, respectively.}\label{Chains}
\end{figure}

We start our analysis by considering time-dependent dephasing models with sinoidal time dependence. 
We focus on two specific situations, called \textit{symmetric} and \textit{non-symmetric} configurations, having in mind the case $N=7$ as a benchmark for the Markovian case~\cite{Elinor}.
In the non-symmetric configuration, the~fifth site is the extraction site, which breaks the inversion symmetry.
In the symmetric configuration, the~extraction site is the $7$th site.
These two configurations are illustrated in Figure~\ref{Chains}.

Chains of different sizes and other choices of extraction sites  breaking inversion symmetry can also be studied, and they present similar behavior to the ones presented here.
In both configurations, the~chain has uniform frequencies $\omega_i = \omega $ and inter-site couplings $\lambda_i = \lambda$. The~injection site is always chosen to be the first site. 
In the symmetric configuration, the~extraction site is on the other tip of the chain, i.e.,~the last site. 

As shown in \cite{Elinor}, {\it Markovian} dephasing-assisted transport manifests only in the non-symmetric configuration.
We investigate what happens when non-Markovian dephasing is introduced in these configurations.
To set a specific scenario, we fix $ \lambda = 0.1\omega$ and $\kappa_{\text{ext}} = \kappa_{\text{inj}} = 0.01\omega$ for all simulations, i.e.,~all frequencies and couplings are set in units of $\omega$.

From an experimental point view, we are in the strong-coupling regime where $\lambda$ is comparable to $\omega$ \cite{strong}. If~one were to consider a typical scenario in a natural system, such as the ones involving exciton transfer complexes, typical parameters would be  $\kappa_{inj}$ ($\kappa_{ext}$) of the order of $ps^{-1}$, on-site gap energies around $\omega_i=10^{4}~\text{cm}^{-1}$ and coupling constants two order of magnitudes weaker $\lambda_i=10^2~\text{cm}^{-1}$. However, our main goal is to illustrate new transport phenomena driven by instances of time dependent-dephasings, which are amenable to quantum simulation~\cite{Liu}.

\subsection{Non-Symmetric~Configuration}

\begin{figure}[h!]
\includegraphics[scale=0.6]{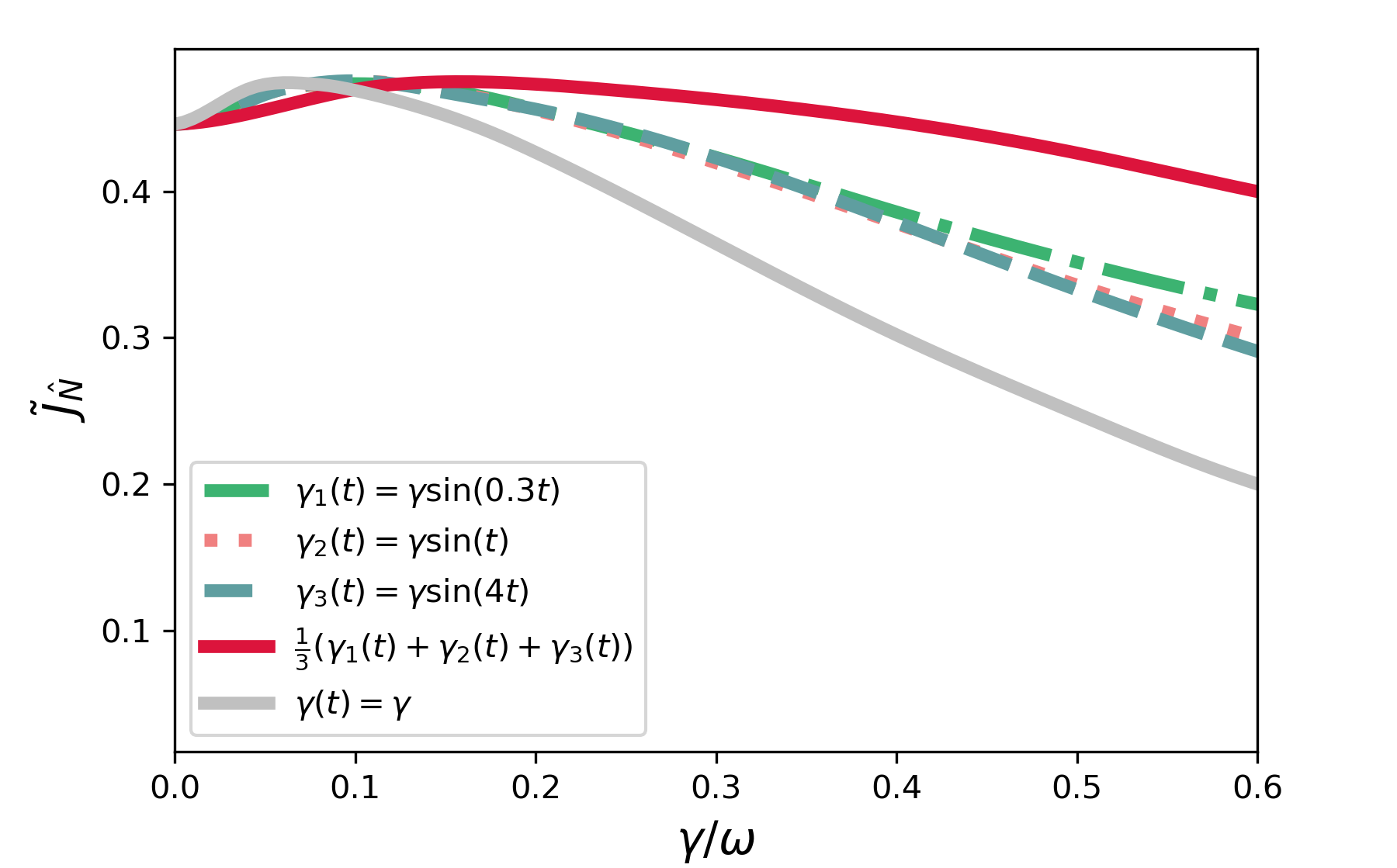}
\label{fig:fig14}
\caption{Non-symmetric configuration -- current $\tilde{J}_{\hat{N}}$ as a function of $\gamma/\omega$ for $\gamma(t) = \gamma \sin(\nu t)$, where $\nu = 0.3, 2, 4$, and~the normalized sum of these functions. The~Markovian case corresponding to $\gamma(t) = \gamma$ is also~plotted.}\label{SUM}
\end{figure}

In Figure~\ref{SUM}, we plot the current $\tilde{J}_{\hat{N}}$ as a function of $\gamma\geq 0$ for the time-dependent dephasing model $\gamma(t) = \gamma \sin(\nu t)$, and~different values of $\nu$ in the non-symmetric configuration. 
We also consider the average of these three sine functions.
This model happens to be non-Markovian for any finite value of the positive constant $\gamma$.
The Markovian case corresponding to $\gamma(t) = \gamma$ is also plotted as a benchmark. The~first point to be noticed is that dephasing-assisted transport manifests in both cases: Markovian and non-Markovian. This corresponds to the first portion of the curves where the current increases with $\gamma$. Regardless of being Markovian or not, there is always an optimal value of $\gamma$ above which dephasing becomes detrimental.
Notwithstanding, we see that the non-Markovian cases are more efficient than their Markovian counterpart for higher dephasing magnitudes $\gamma$.


\begin{figure}[h!]
\includegraphics[scale=0.6]{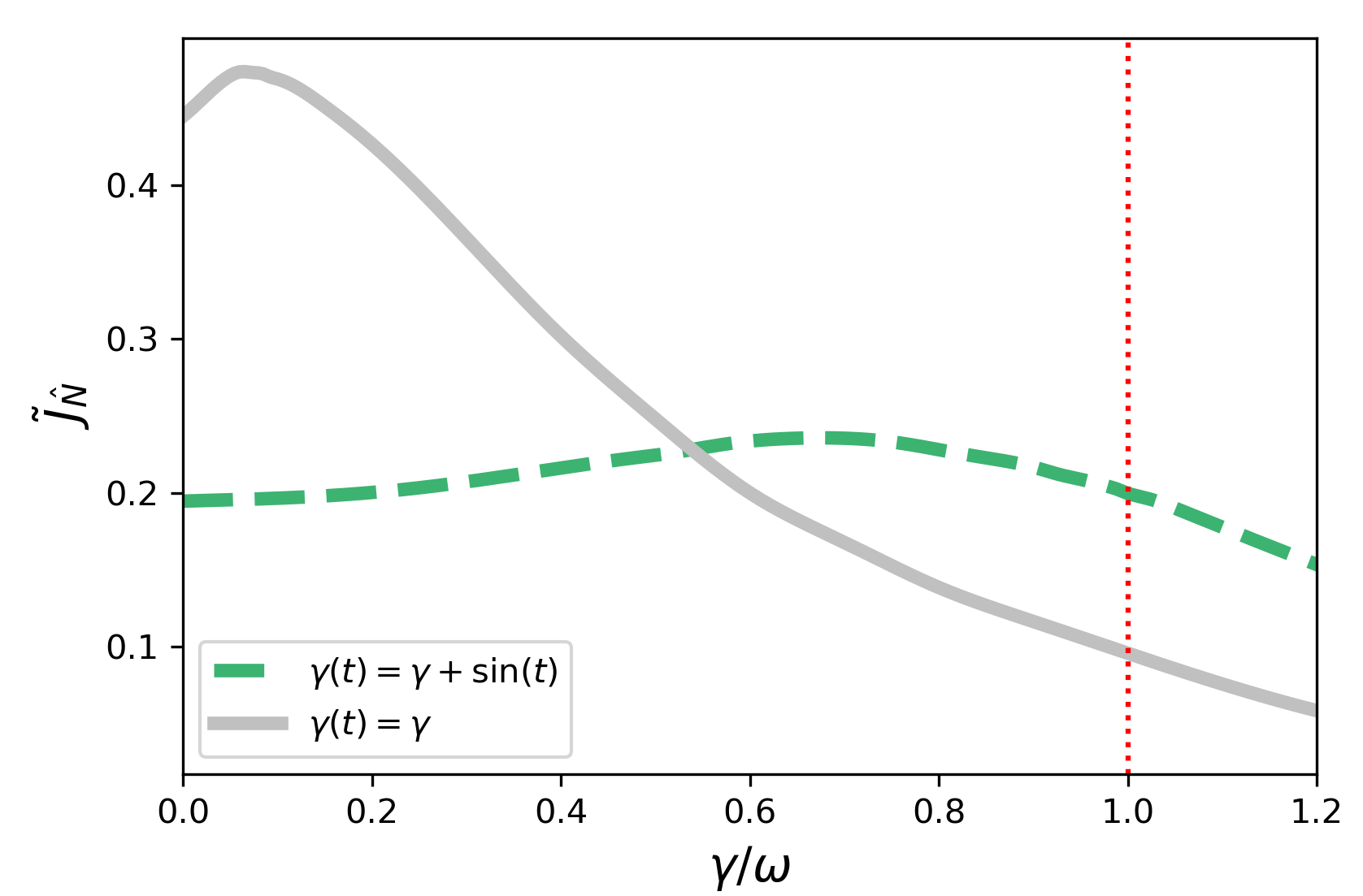}
\label{fig:fig14}
\caption{Non-symmetric configuration -- current $\tilde{J}_{\hat{N}}$ as a function of $\gamma/\omega$ for  $\gamma(t) = \gamma + \gamma_0\sin( t)$. The~dotted red vertical line corresponds to $\gamma = 1$. The~system is decreasingly non-Markovian in the interval $0 < \gamma < 1$. For~$\gamma \geq 1$, the~system is Markovian since we have $\gamma(t) \geq 0$ for all $t$. The~gray curve corresponds  to $\gamma(t) = \gamma$.}\label{AllSINt}
\end{figure}

In Figure~\ref{AllSINt}, we consider the current $\tilde{J}_{\hat{N}}$ for another model, for~which $\gamma(t) = \gamma + \gamma_0\sin( t)$ in the non-symmetric configuration, with~$\gamma_0 = 1$. Now we have a transition from non-Markovian to Markovian depending on a physical parameter, i.e.,~the resulting dynamics is non-Markovian for $0<\gamma< 1$. 
One can also see the time-independent Markovian benchmark in the same plot. 
As a glimpse of how rich the transport scenario is in the presence of time-dependent dephasing, this model does not present efficiency enhancement by dephasing.  

Compared to the Markovian case for $\gamma=0$, i.e.,~closed system dynamics, the~case with $\gamma(t) = \gamma + \sin( t)$ is always less efficient. 
This is in clear contrast to the model considered before. 
However, the~present model shows an interesting non-monotonic behavior with $\gamma$, and~it also turns out to be more efficient than the time-independent Markovian counterpart for higher values of $\gamma$.

\unskip

\subsection{Symmetric~Configuration}

\begin{figure}[h!]
\includegraphics[scale=0.65]{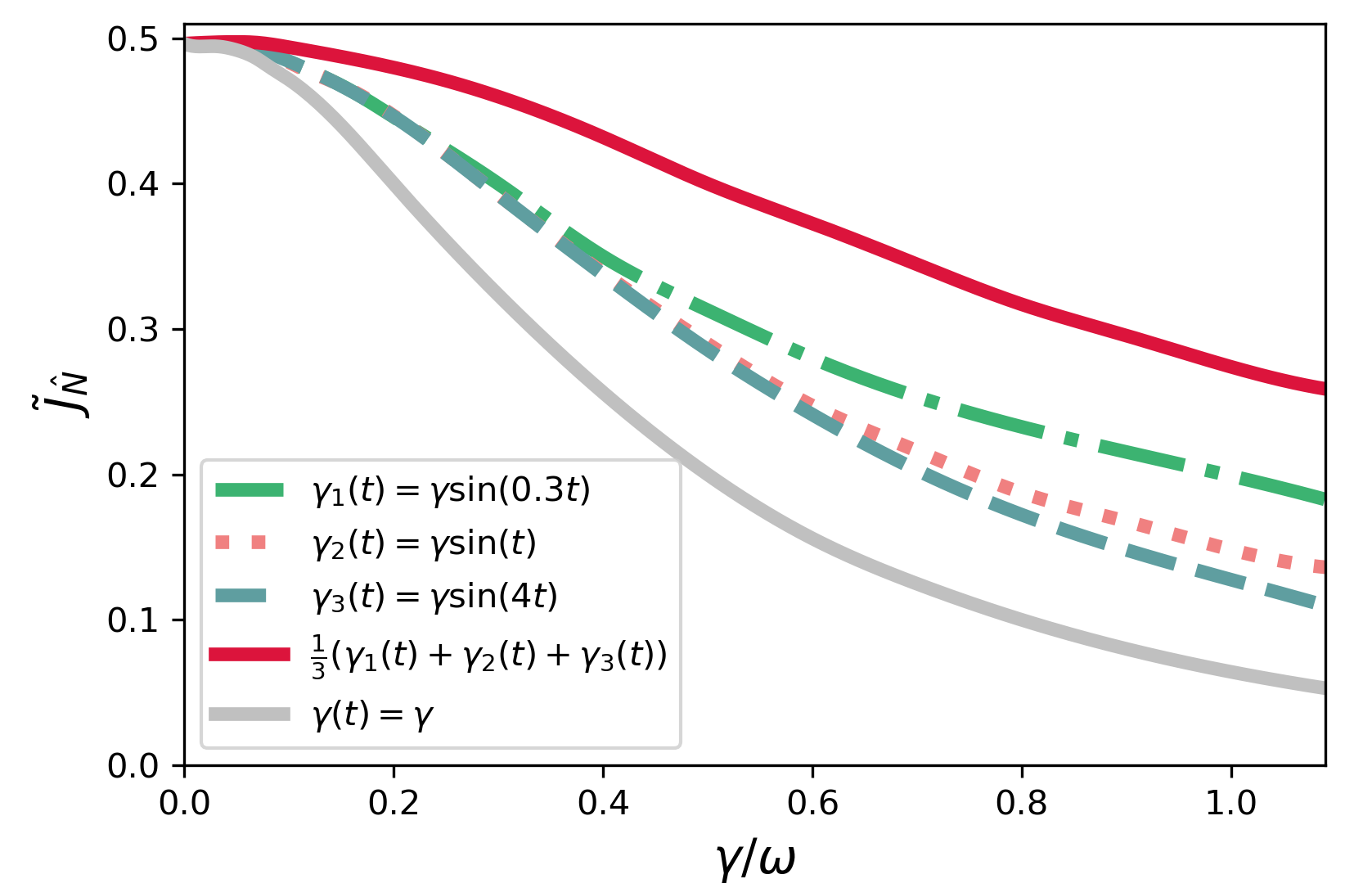}
\label{fig:fig14}
\caption{Symmetric configuration -- current $\tilde{J}_{\hat{N}}$ as a function of $\gamma/\omega$ for $\gamma(t) = \gamma \sin(\nu t)$, where $\nu = 0.3, 2, 4$, and~the normalized sum of these functions. The~Markovian case corresponding to $\gamma(t) = \gamma$ is also~plotted.}\label{SIMSUM}
\end{figure}

Now, we focus on the symmetric configuration, and again plot the current $\tilde{J}_{\hat{N}}$ as a function of $\gamma$ for the model  $\gamma(t) = \gamma \sin(\nu t)$.
The Markovian case corresponding to $\gamma(t) = \gamma$ is also plotted and shows a monotonic behavior as $\gamma$ is increased. 
In other words, there is no Markovian dephasing-assisted transport in the symmetric case, which is in~agreement with \cite{Elinor}. 
For each non-Markovian curve shown in Figure~\ref{SIMSUM}, we also have that non-Markovian dephasing-assisted transport does not manifest, as~the maximum current is reached for $\gamma = 0$. 
Nevertheless, it is remarkable to see that the non-Markovian cases become, once again, more efficient than their Markovian counterparts as $\gamma$ is~increased.

\begin{figure}[h!]
\includegraphics[scale=0.65]{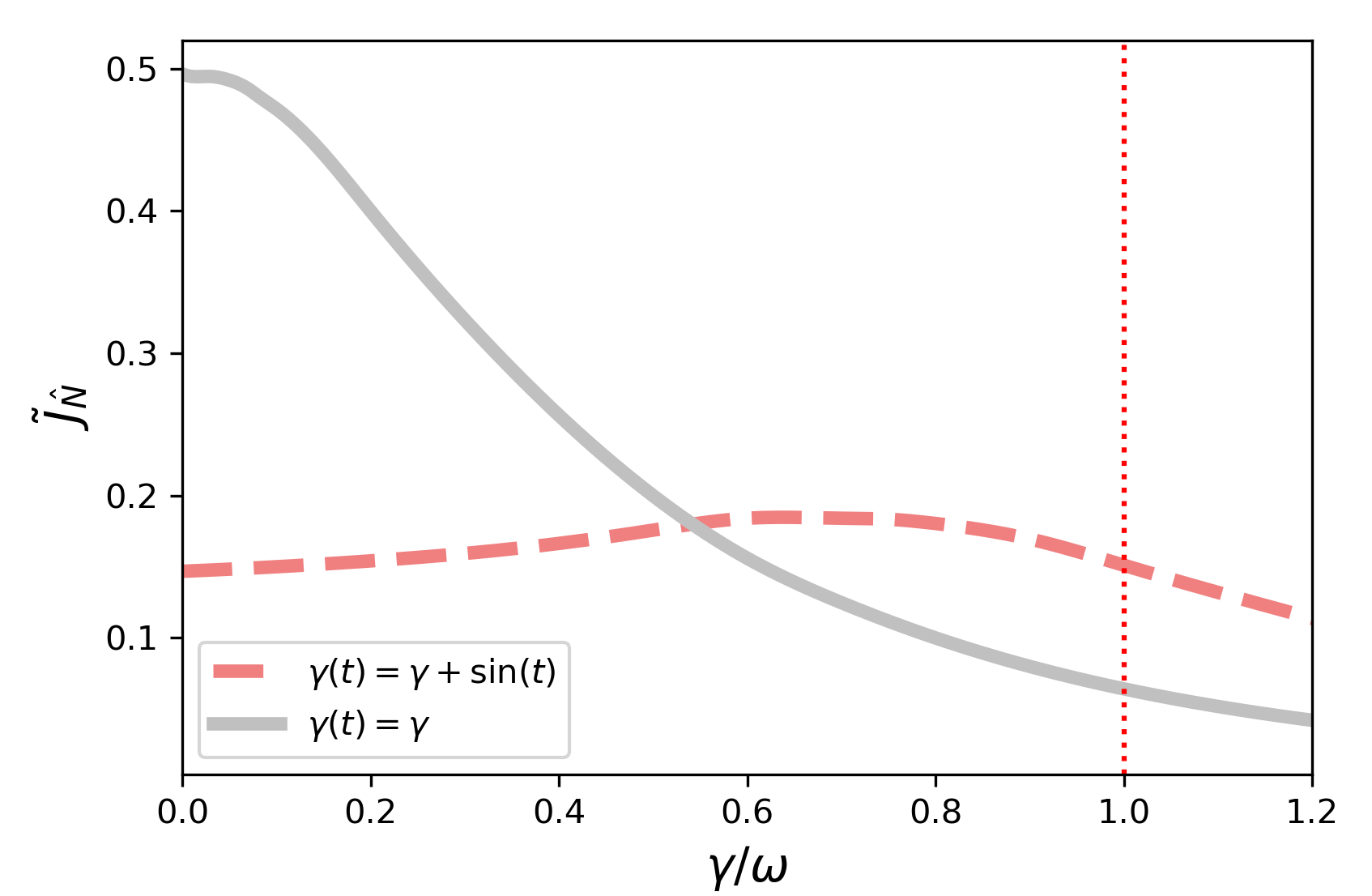}
\label{fig:fig14}
\caption{Symmetric configuration -- current $\tilde{J}_{\hat{N}}$  as a function of $\gamma/\omega$ for  $\gamma(t) = \gamma + \gamma_0\sin(t)$. The~dotted red vertical line corresponds to $\gamma = 1$. The~system is decreasingly non-Markovian in the interval $0 < \gamma < 1$. For~$\gamma \geq 1$, the~system is Markovian since we have $\gamma(t) \geq 0$. The~gray curve corresponds  to $\gamma(t) = \gamma$.}\label{SIMSINt}
\end{figure}

In Figure~\ref{SIMSINt}, we consider once again the model given by $\gamma(t)=\gamma + \gamma_0\sin(\nu t)$, with~$\gamma_0 = 1$ in the symmetric configuration. As~we can see, a~non-monotonic behavior is also observed in this case, and, by~comparing it to the the benchmark, we see that it can also help efficiency regardless of being Markovian $(\gamma<1)$ or not $(\gamma\geq 1)$. 

\unskip


\subsection{Spread of Occupations and~Efficiency}

\begin{figure}[h!]
\centering 
\subfloat[]{\includegraphics[width = 1.6in]{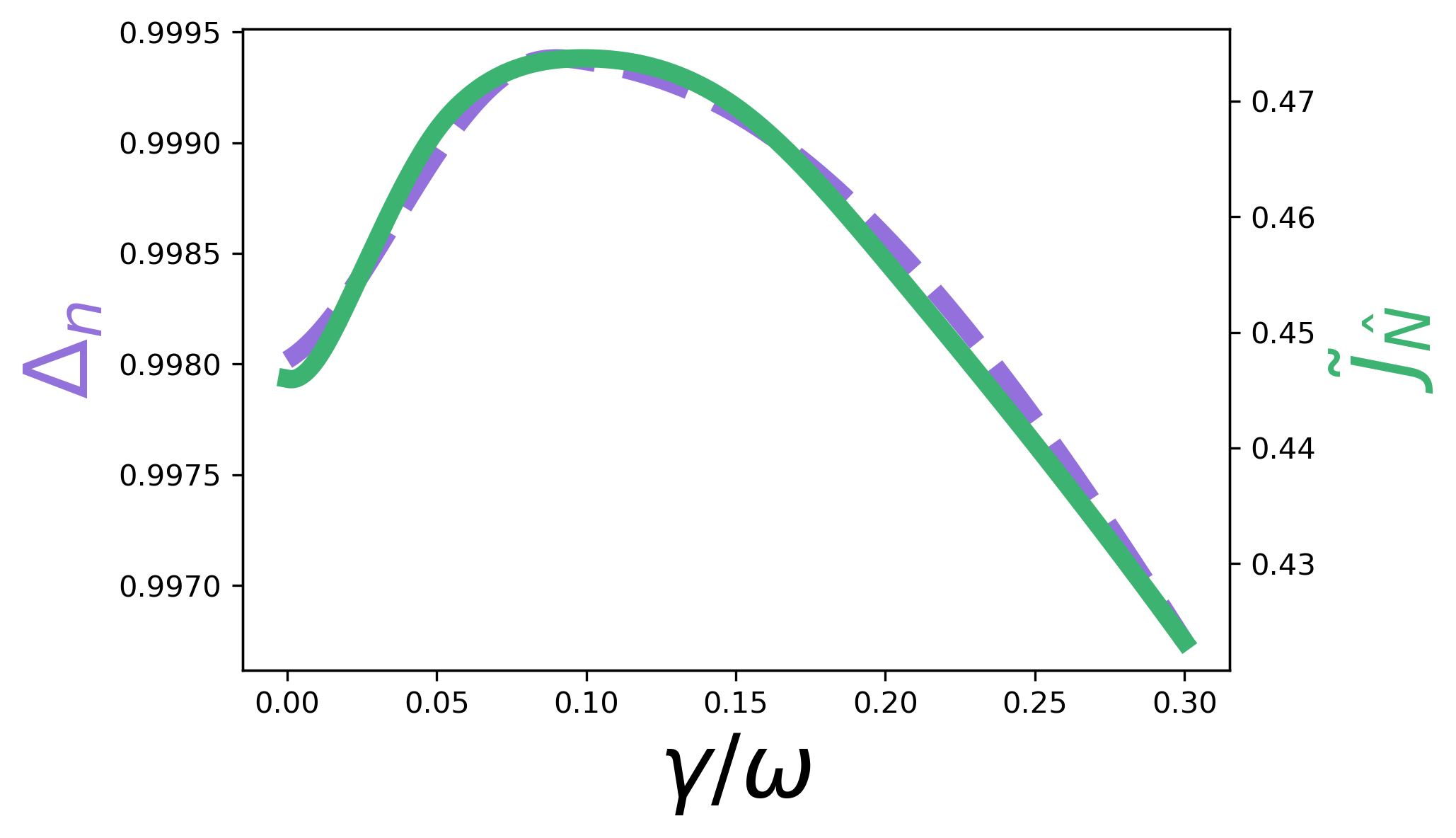}} 
\subfloat[]{\includegraphics[width = 1.6in]{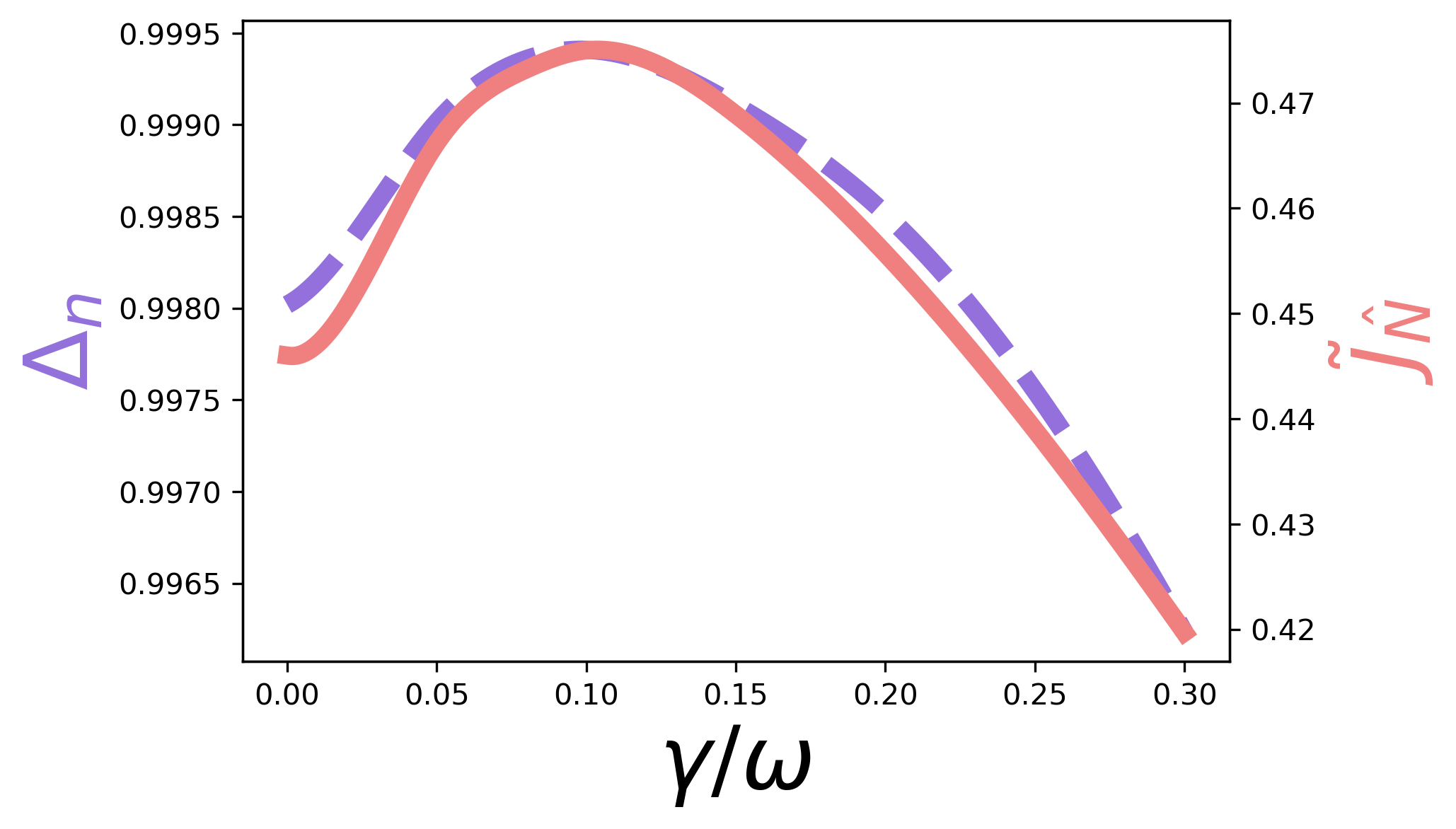}}
\subfloat[]{\includegraphics[width = 1.6in]{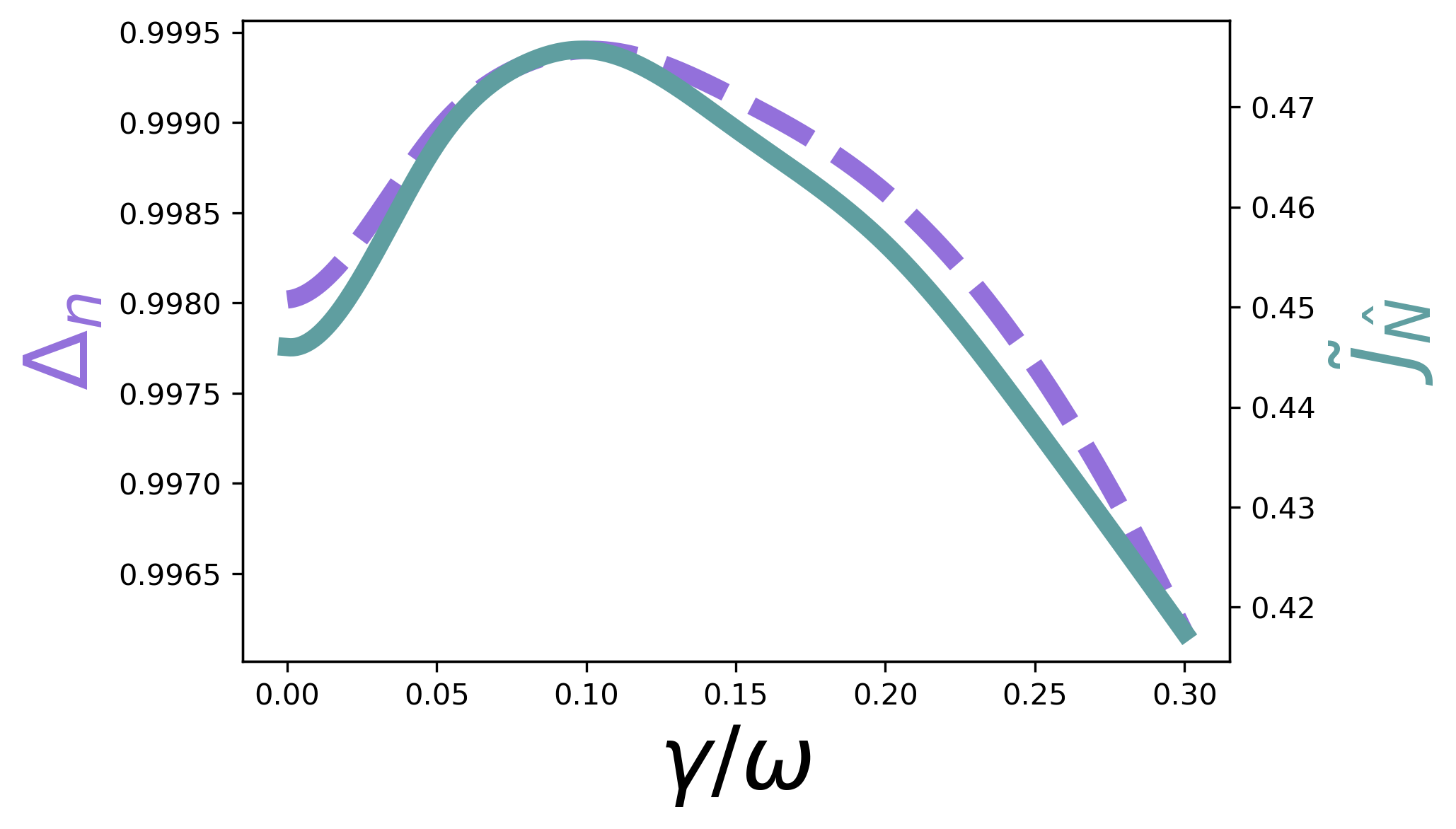}}
\subfloat[]{\includegraphics[width = 1.6in]{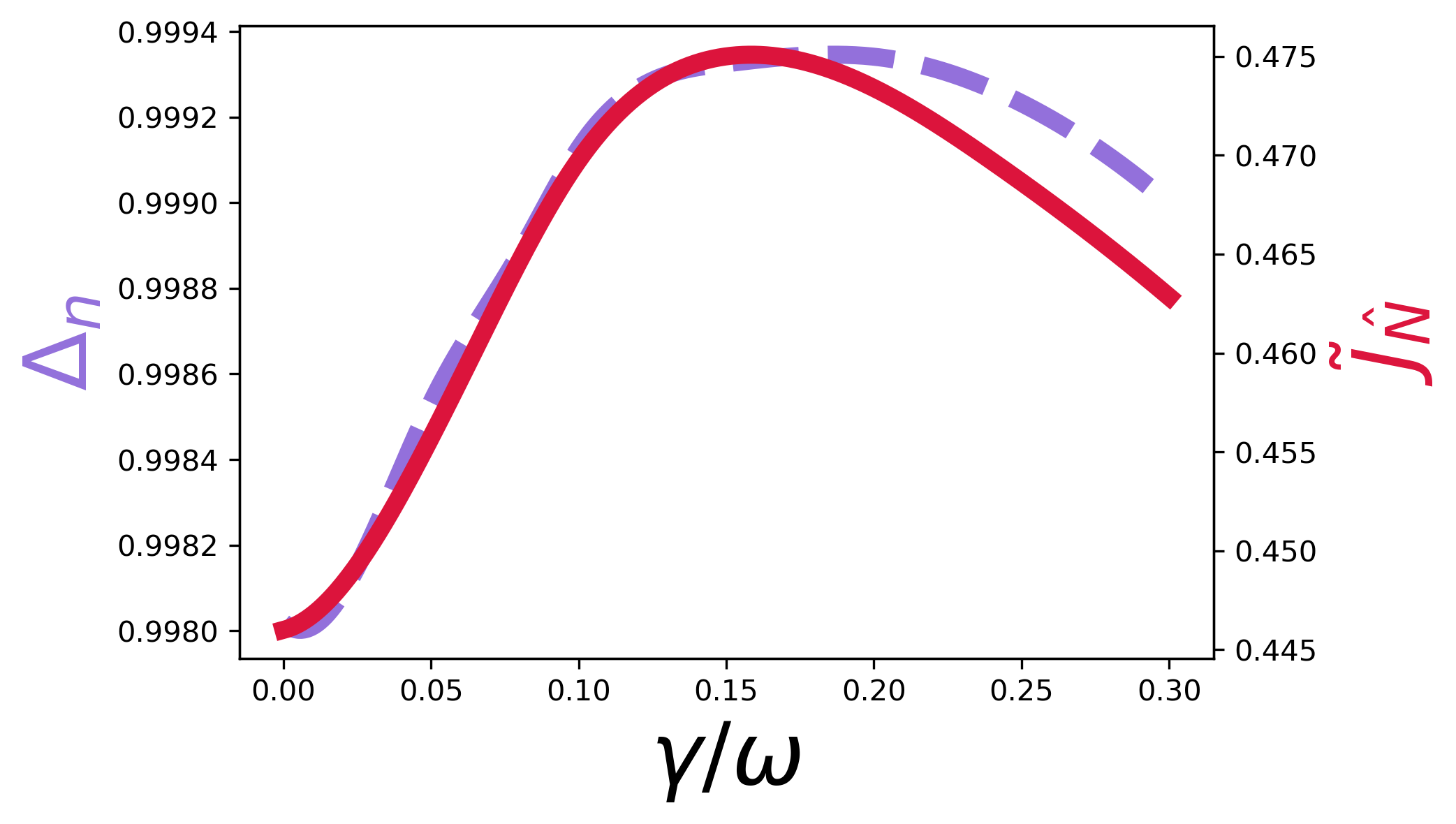}}

\subfloat[]{\includegraphics[width = 1.6in]{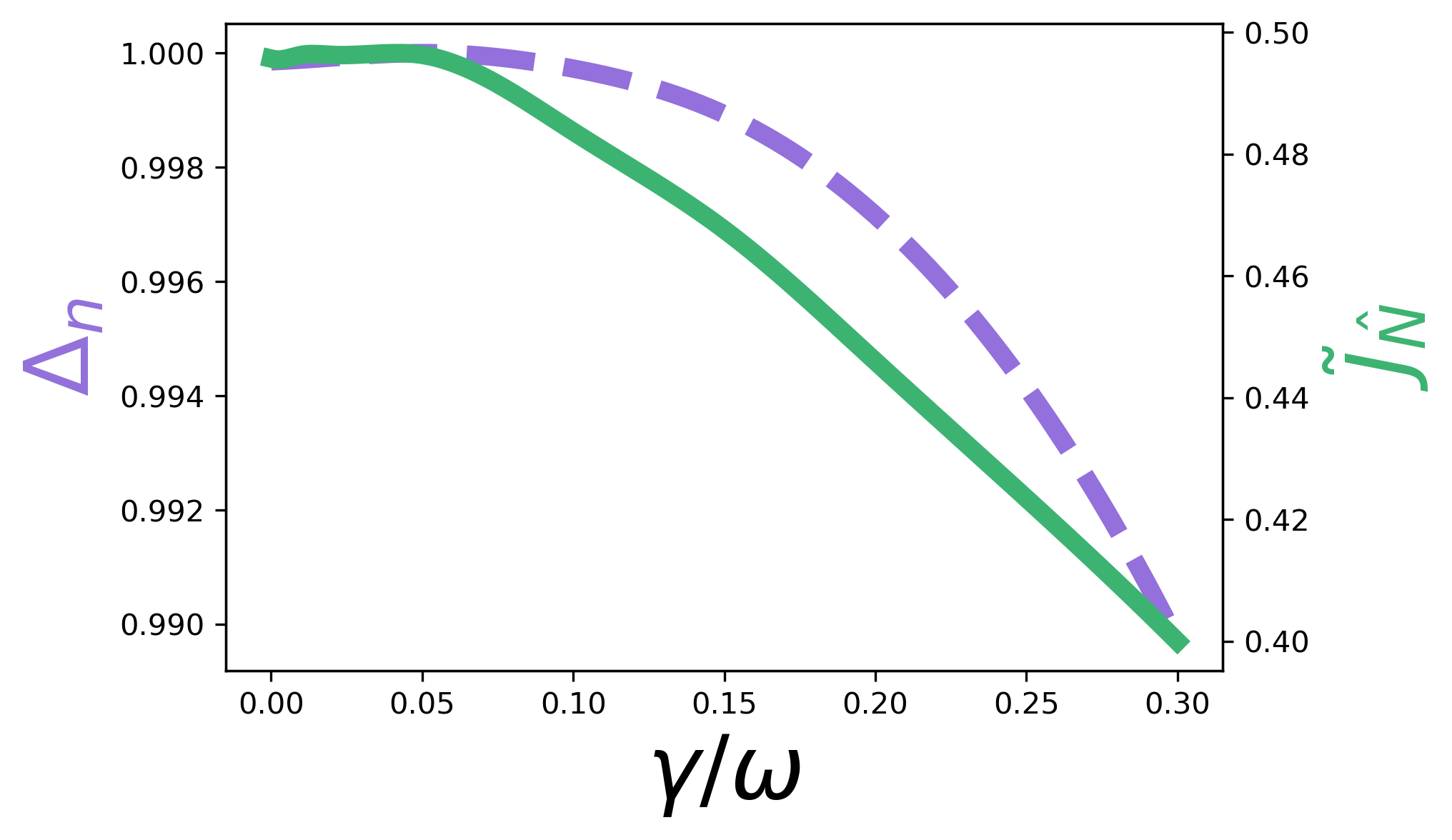}} 
\subfloat[]{\includegraphics[width = 1.6in]{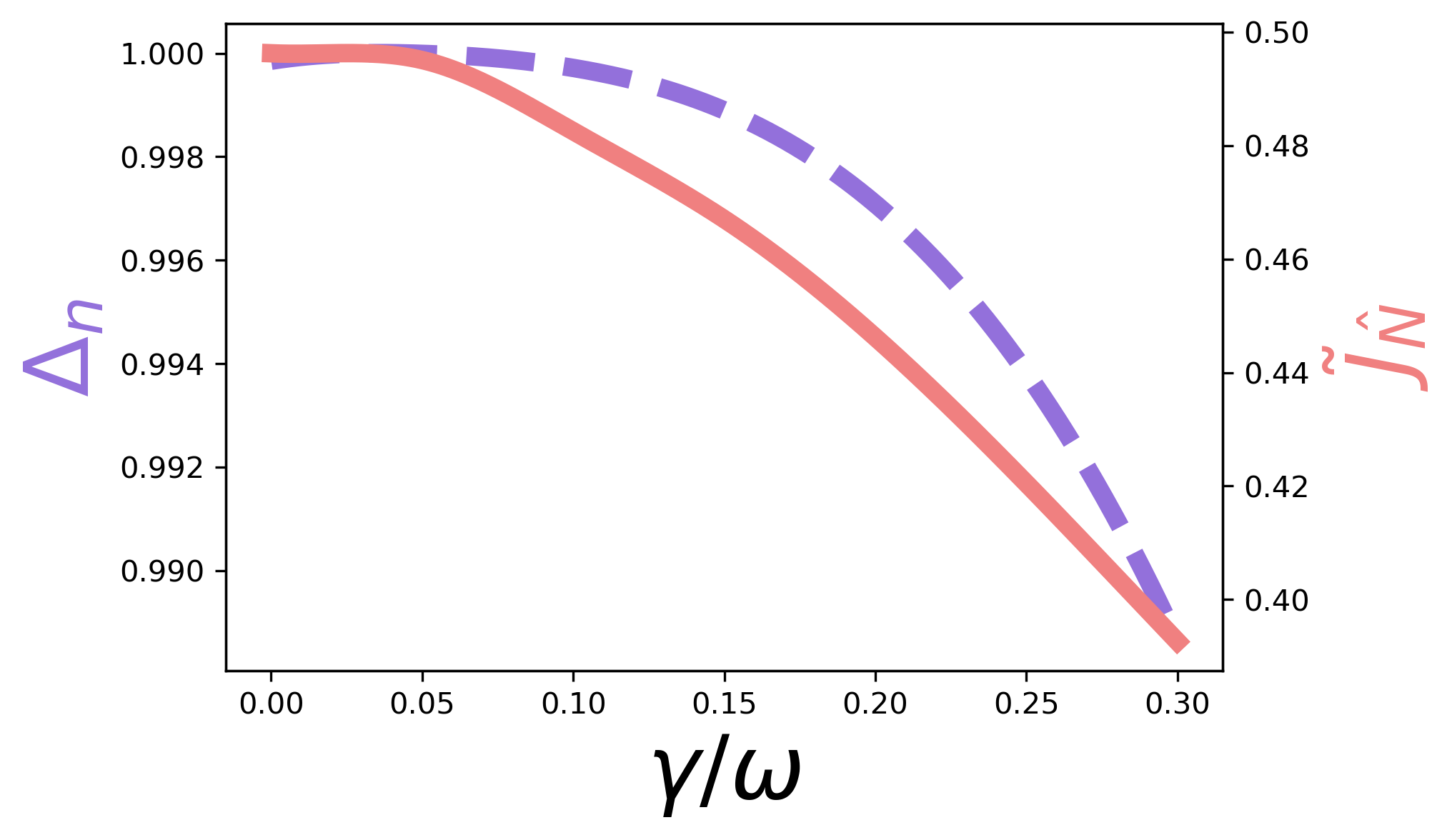}}
\subfloat[]{\includegraphics[width = 1.6in]{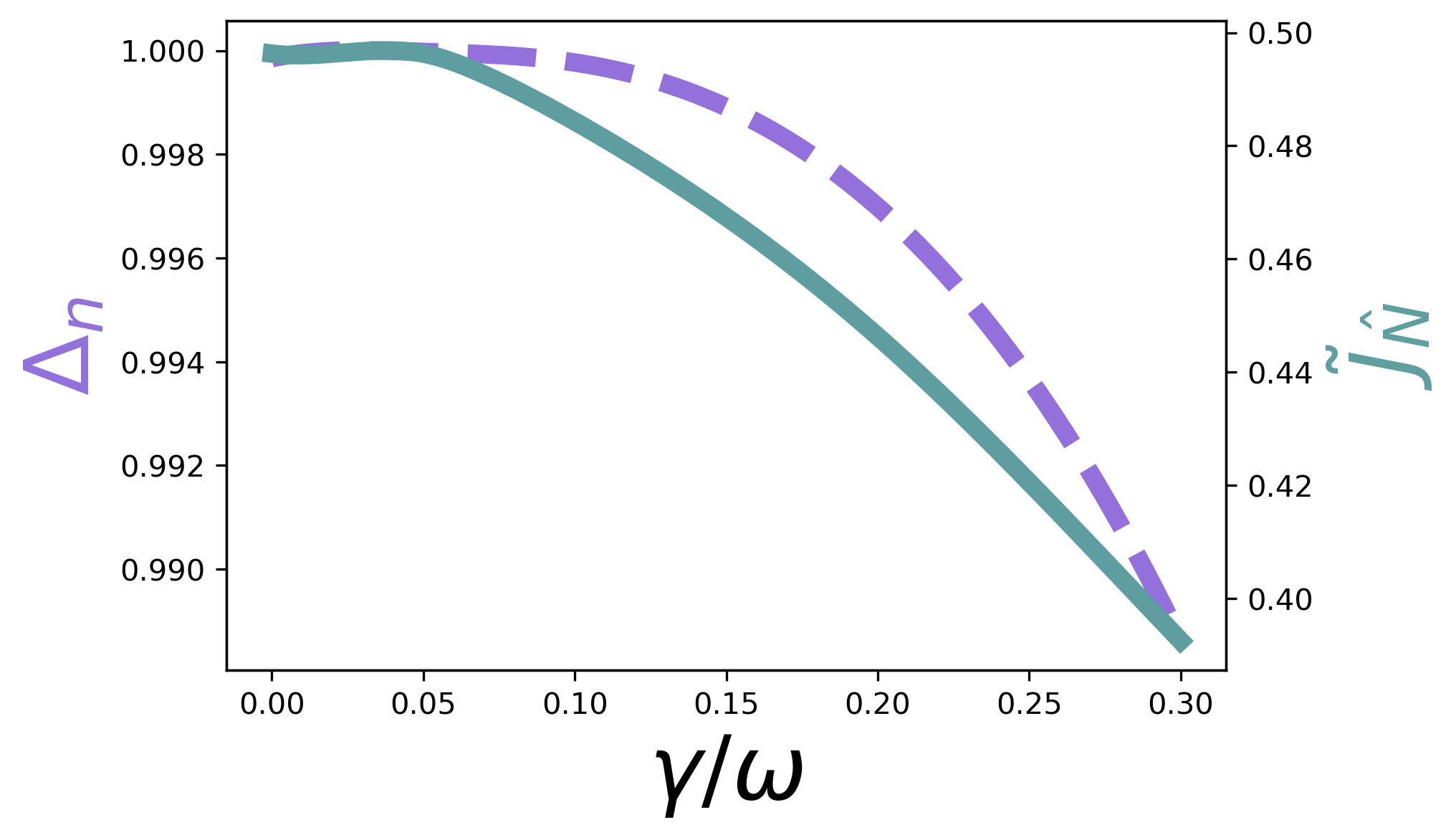}}
\subfloat[]{\includegraphics[width = 1.6in]{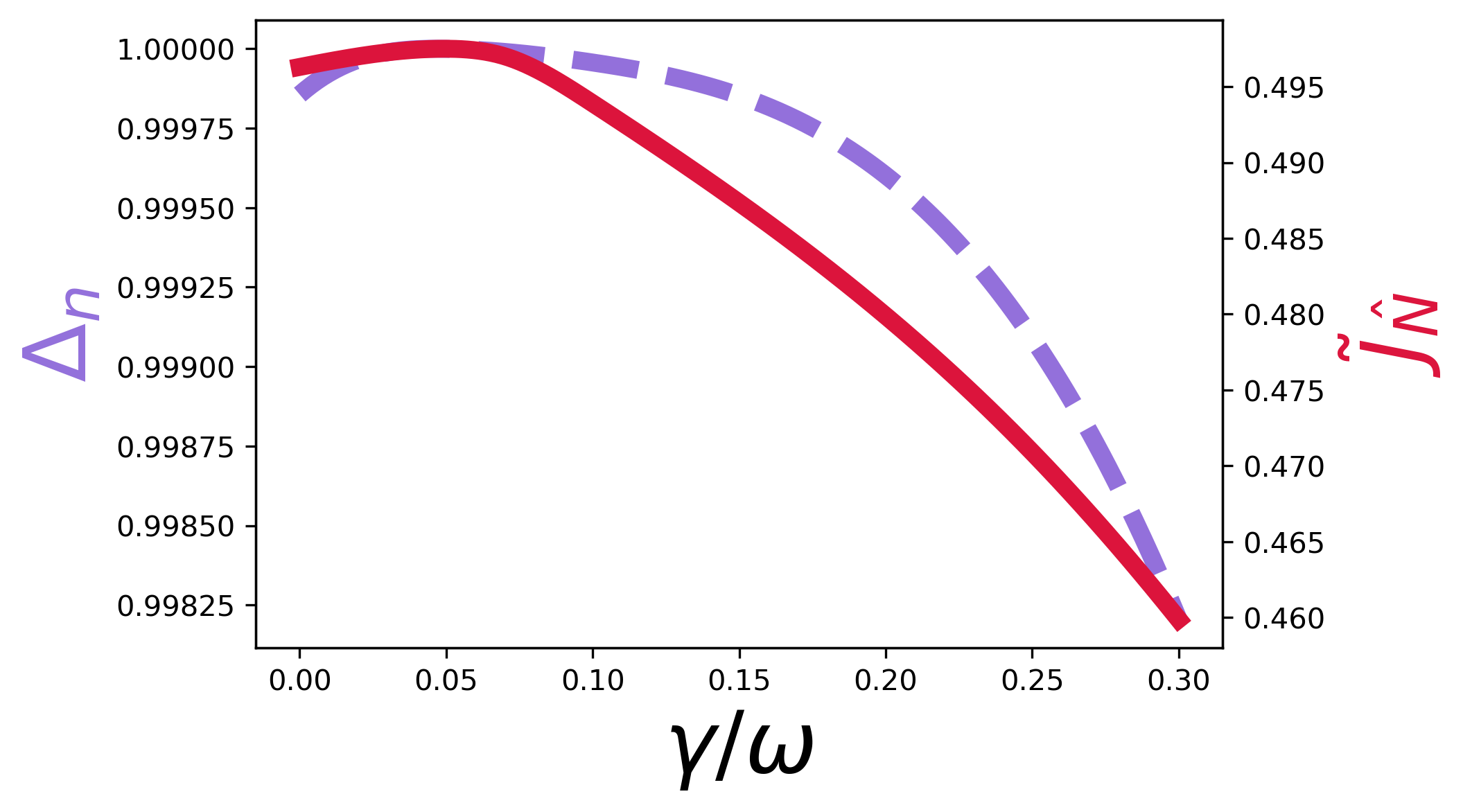}} 

\caption{ $\Delta_n$ (dashed purple line) and the current $\tilde{J}_{\hat{N}}$ (solid line) as a function of $\gamma/\omega$ for each non-Markovian evolution, corresponding to the $\gamma(t)$ in the caption of each figure. The~plots in figures (\textbf{a})--(\textbf{d}) correspond to the non-symmetric configuration, while the plots in Figures (\textbf{e})--(\textbf{h}) correspond to the symmetric~configuration. (\textbf{a}) $\gamma_1(t) = \gamma \sin(0.3 t)$. (\textbf{b})~$\gamma_2(t) = \gamma \sin( t)$. (\textbf{c}) $\gamma_3(t) = \gamma \sin(4 t)$. (\textbf{d}) $\gamma(t) = \frac{1}{3}(\gamma_1(t) + \gamma_2(t) + \gamma_3 (t))$. (\textbf{e}) $\gamma_1(t) = \gamma \sin(0.3 t)$. (\textbf{f}) $\gamma_2(t) = \gamma \sin( t)$. (\textbf{g}) $\gamma_3(t) = \gamma \sin(4 t)$. (\textbf{h}) $\gamma(t) =  \frac{1}{3}(\gamma_1(t) + \gamma_2(t) + \gamma_3 (t))$.}\label{DeltaN}
\label{some example}
\end{figure}
\begin{figure}[h!]

\subfloat[]{\includegraphics[width = 2.8in]{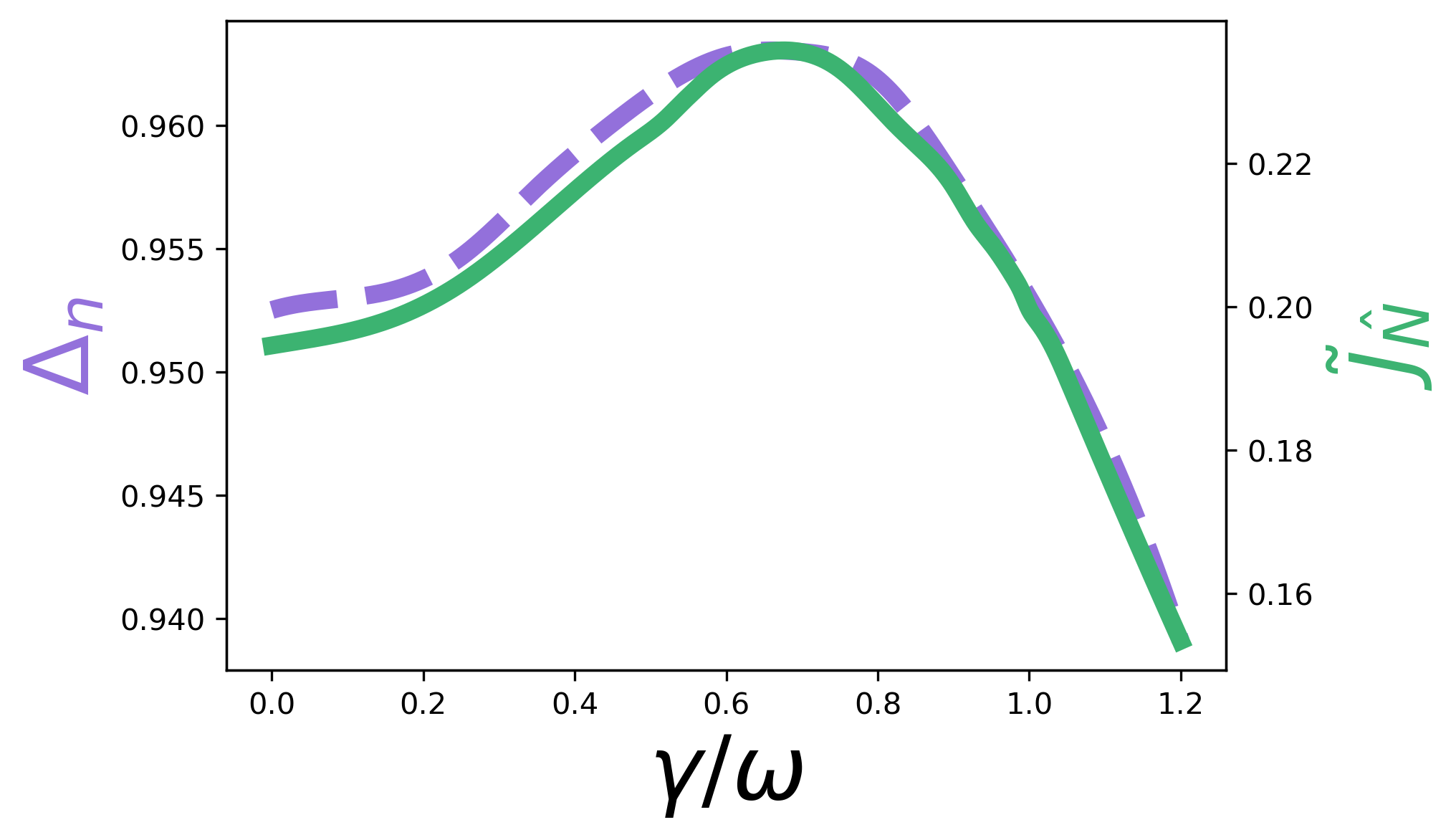}} 

\subfloat[]{\includegraphics[width = 2.8in]{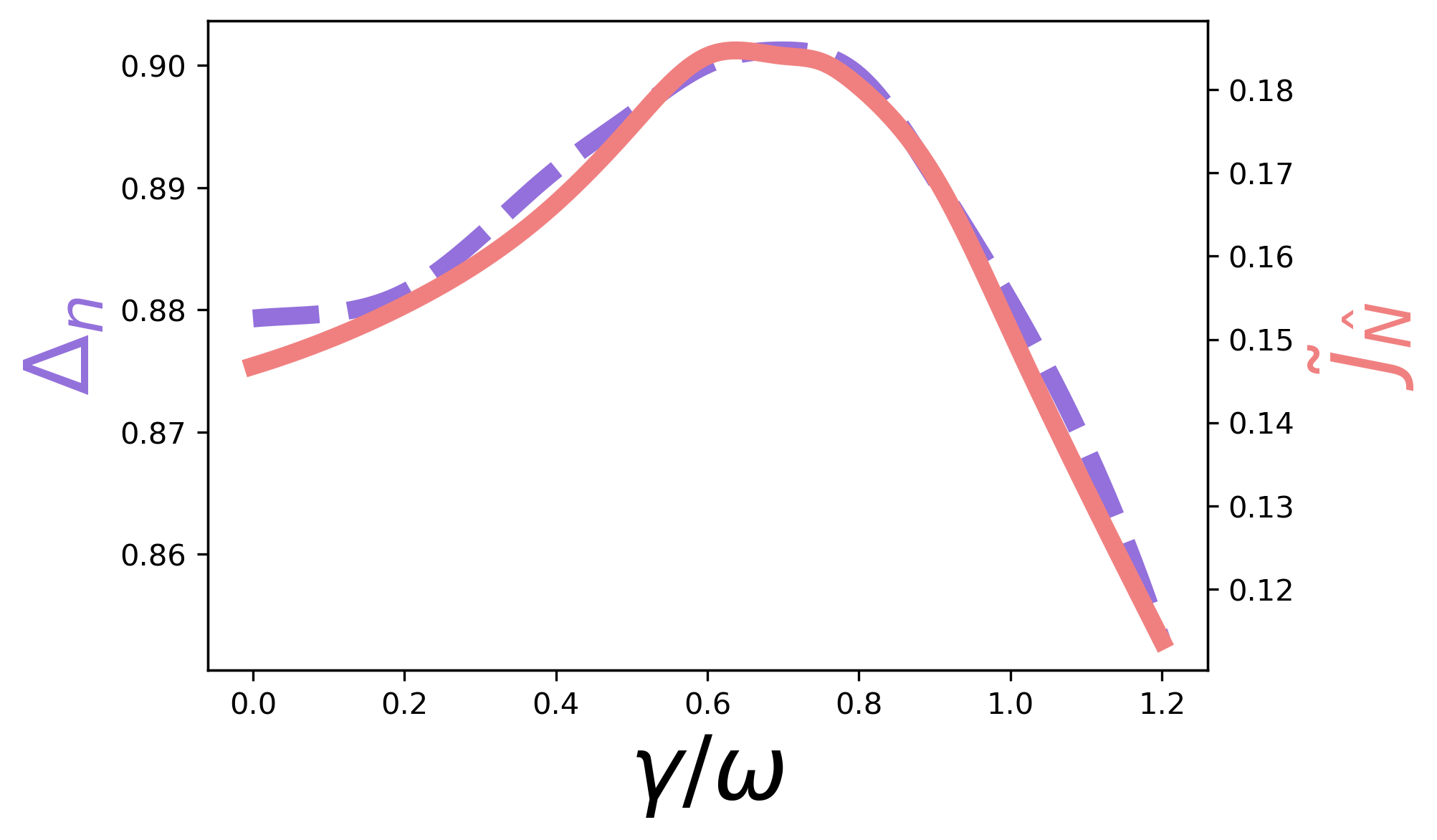}}
\caption{ $\Delta_n$  (dashed purple line) and  the current $\tilde{J}_{\hat{N}}$ (green/coral solid lines) as a function of $\gamma/\omega$, for~$\gamma(t) = \gamma + \sin( t)$ for the non-symmetric and symmetric~configurations. (\textbf{a}) Non-symmetric config. (\textbf{b}) Symmetric~config.}\label{DeltaNGamma}
\label{some example}
\end{figure}

Next, we seek to analyse how the spread of occupations correlates with the current maximum in the time-dependent dephasing scenarios presented above.
We consider the spread of occupations $\Delta_n$ \cite{Elinor}, with~$n_i \equiv p_i(\infty)$,
\begin{equation}
\Delta_n = 1 - \left(\frac{1}{N} \sum_i n_i - n_{\text{k}}\right)^2,
\end{equation}
where $n_k$ is the population of the extraction site $k$, $n_k \equiv p_k(\infty)$.
The maximum of this quantity is associated with a minimum spread of the occupations.
A correlation between the maximal of $\Delta_n$ and the maximum of the current is verified in \cite{Elinor} for several time-independent Markovian cases.
Here, we certify that, for~the time-dependent non-Markovian cases studied above, the~same tendency is verified: $\Delta_n$ is maximum when the current is maximum, as~shown in Figures~\ref{DeltaN} and \ref{DeltaNGamma}, which shows plots of $\Delta_n$ and the current as a function of $\gamma$.
These results suggest that this quantity is an indicator of optimal transport scenarios in the more general time-dependent and non-Markovian dephasing~picture.

\nointerlineskip

\vspace{-6pt}

\unskip


\section{Example: Controlled Quantum~System}\label{example}

In the scope of controlled quantum systems, time-dependent dephasing, including non-Markovian evolutions, can be introduced and externally controlled~\cite{Souza,paulo,fred,mauro, exp}.
In a recent work~\cite{Ancheyta}, for~example, it is shown that the dynamics of a driven quantum harmonic oscillator subject to non-dissipative noise is formally equivalent to single-particle dynamics in a dynamically-disordered photonic network, and~it is shown that non-Markovianity can be advantageous in the noise-assisted transport phenomenon.

In quantum technologies, one can induce non-Markovianity through the use of controlled auxiliary systems. 
Specifically, a~model in the context of nuclear magnetic resonance (NMR) experiments, where a Ising-like interaction takes place between two spin 1/2 systems, is studied in \cite{Souza}.
One of these two-level systems is considered to be the system of interest, and~the other is seen as part of the environment, providing a structured bath.

The strength of the coupling between the system and the environment is given by the parameter $J$, and~$\theta$ is a parameter that gives the state in which the environment is initialized, before~the interaction.
It turns out that the parameters $J$ and $\theta$ are controllable in the NMR experimental realization.
In particular, the~following superoperator can be engineered for any site $i$ of the chain  \citep{Souza}
\begin{equation}\label{Li}
\mathcal{L}_i\rho = \frac{1}{2}\gamma_i(t)(\sigma_i^z\rho\sigma_i^z-\rho)-is_i(t)[\sigma_i^z,\rho],
\end{equation}
where
\begin{equation}\label{gammat}
\gamma_i(t)= \gamma_i+ \frac{\pi J \sin^2(2\theta)\sin(2 \pi J t)}{3+2\cos(4\theta)\sin^2(\pi J t)+\cos(2 \pi J t )},
\end{equation}
is a time-dependent dephasing rate, and~\begin{equation}\label{st}
s_i(t)=\frac{2 \pi J \cos(2\theta)}{3+2\cos(4\theta)\sin^2(\pi J t)+\cos(2 \pi J t )},
\end{equation}
is an environment-induced time-dependent energy shift. 
As mentioned before, in~\linebreak \mbox{Equations~(\ref{gammat}) and (\ref{st})}, $J$ and $\theta$ are fully controlled~parameters. 

\begin{figure}[h!]
\includegraphics[scale=0.6]{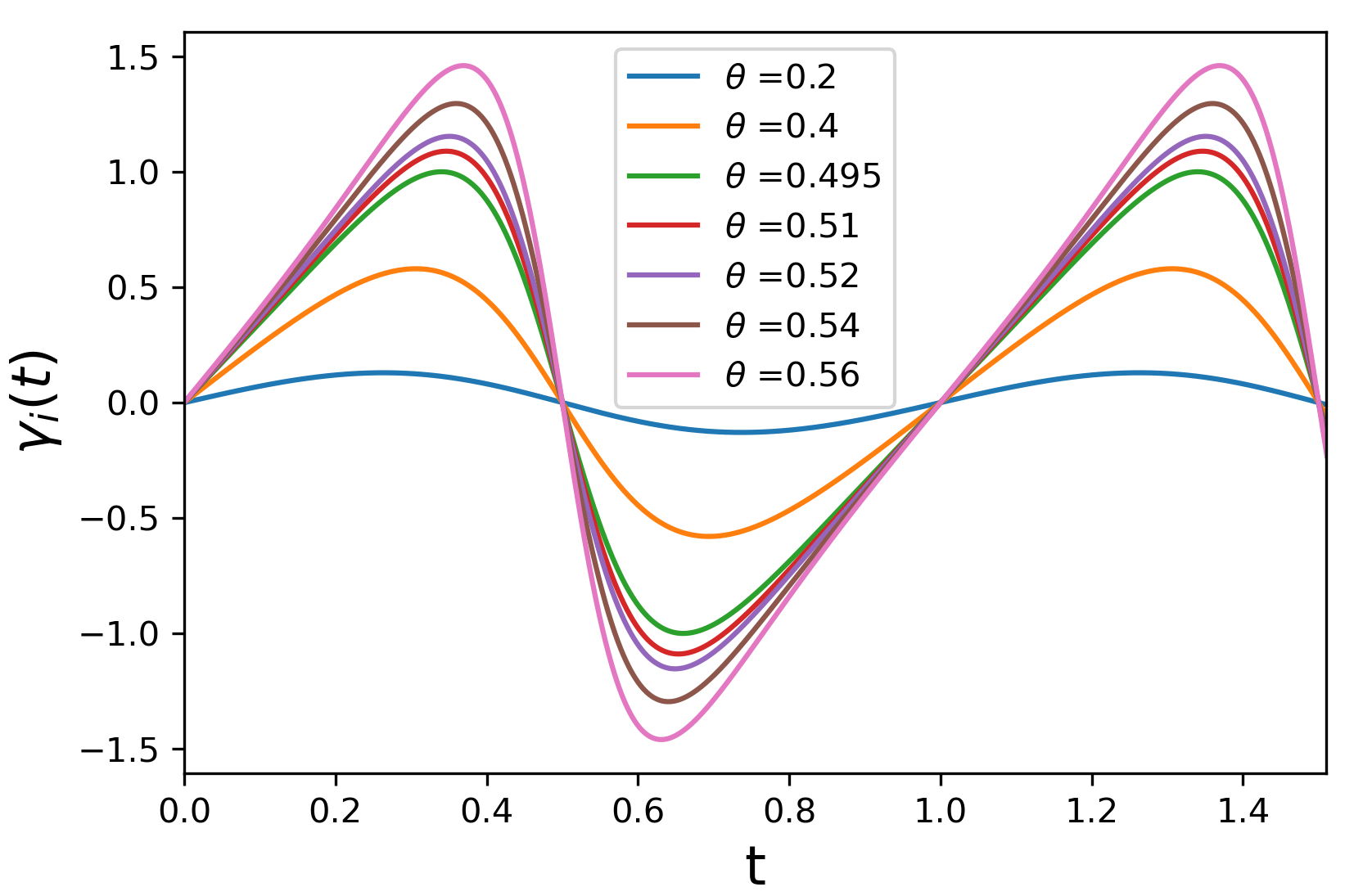}
\label{fig:fig14}
\caption{$\gamma(t)$ in Equation \eqref{gammat} with $\gamma = 0$ and $J = 1$ and different values of $\theta$. 
}\label{Gammas}
\end{figure}

First, it is worth studying the behavior of the function $\gamma_i(t)$ in Equation \eqref{gammat}, which is a periodic function satisfying the condition for a completely positive evolution for any value of $\gamma_i \ge 0$ as discussed in Section~\ref{scnm}.
In all the plots in this section, we will consider $J=1$. 
In Figure~\ref{Gammas}, we plot $\gamma_i(t)$ for several values of the parameter $\theta$ while keeping $\gamma _i= 0$.

We can see that, for~$\gamma_i = 0$ the system is always non-Markovian, and~its non-Markovianity increases as $\theta$ increases in the interval $[0, \pi/2]$.
For simplicity, we assume that $\gamma_i = \gamma$ so that all $\gamma_i(t)$ are the same.
Thus, the~whole chain will be subjected to the following total master equation, which also takes into account the coupling to the incoherent energy sources responsible for injection and extraction of energy, as~described~before,
\begin{equation}\label{MasterControlled}
\dot{ \rho} =-i [H,\rho] +  \sum_i^N  \frac{1}{2} \gamma_i(t) (\sigma_i^z \rho \sigma_i^z - \rho) - i s_i(t)[\sigma_i^z, \rho]  +  \mathcal{L}_{\text{inj}}\rho + \mathcal{L}_{\text{ext}}\rho.
\end{equation}

\subsection{Non-Symmetric~Configuration}

\begin{figure}[h!]
\includegraphics[scale=0.6]{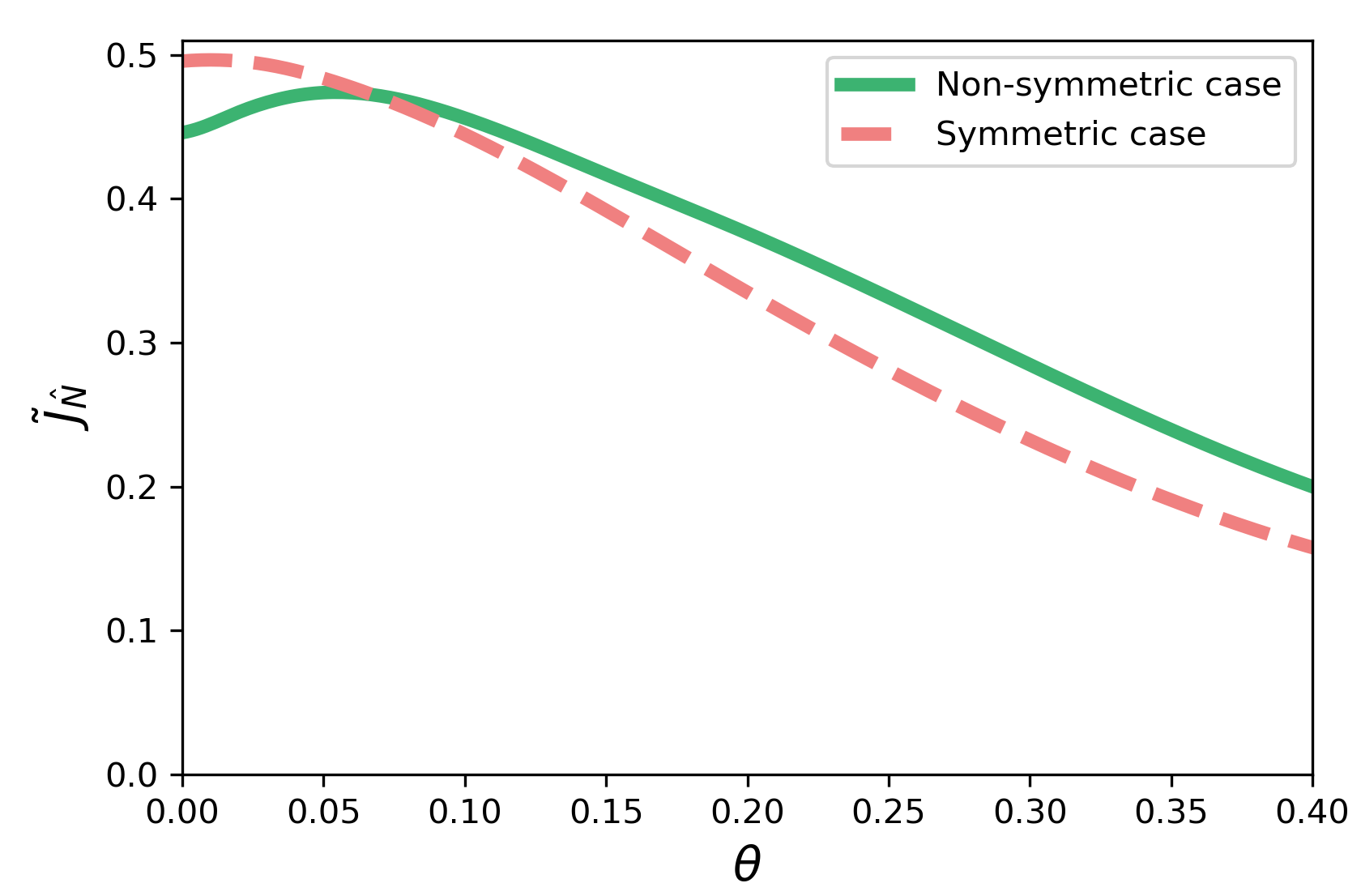}
\label{fig:fig14}
\caption{Current $\tilde{J}_{\hat{N}}$ as a function of $\theta$ (rad) for $\gamma = 0$. The~solid green line represents the non-symmetric case, while the dashed coral line represents the symmetric~one.}\label{DAT}
\end{figure}

We now focus on the model with time-dependent dephasing as described by \linebreak Equations~\eqref{gammat}--\eqref{MasterControlled}.
First, we consider $\gamma = 0$ in Equation \eqref{gammat}.
In Figure~\ref{DAT}, we have the current $\tilde{J}_{\hat{N}}$ plotted as a function of $\theta$---see the solid green line.
We see that non-Markovian dephasing-assisted transport happens, as~the maximum of the current is associated with a value $\theta \neq 0$.
This is similar to the behavior discussed before and illustrated in \mbox{Figure~\ref{SUM}}, but~here $\theta$ is the parameter controlling the non-Markovian dephasing.
In other words, by~increasing $\theta$ in the range shown, one increases the presence of non-Markovian dephasing in the system's~evolution.

%

\begin{figure}[h!]
\includegraphics[scale=0.6]{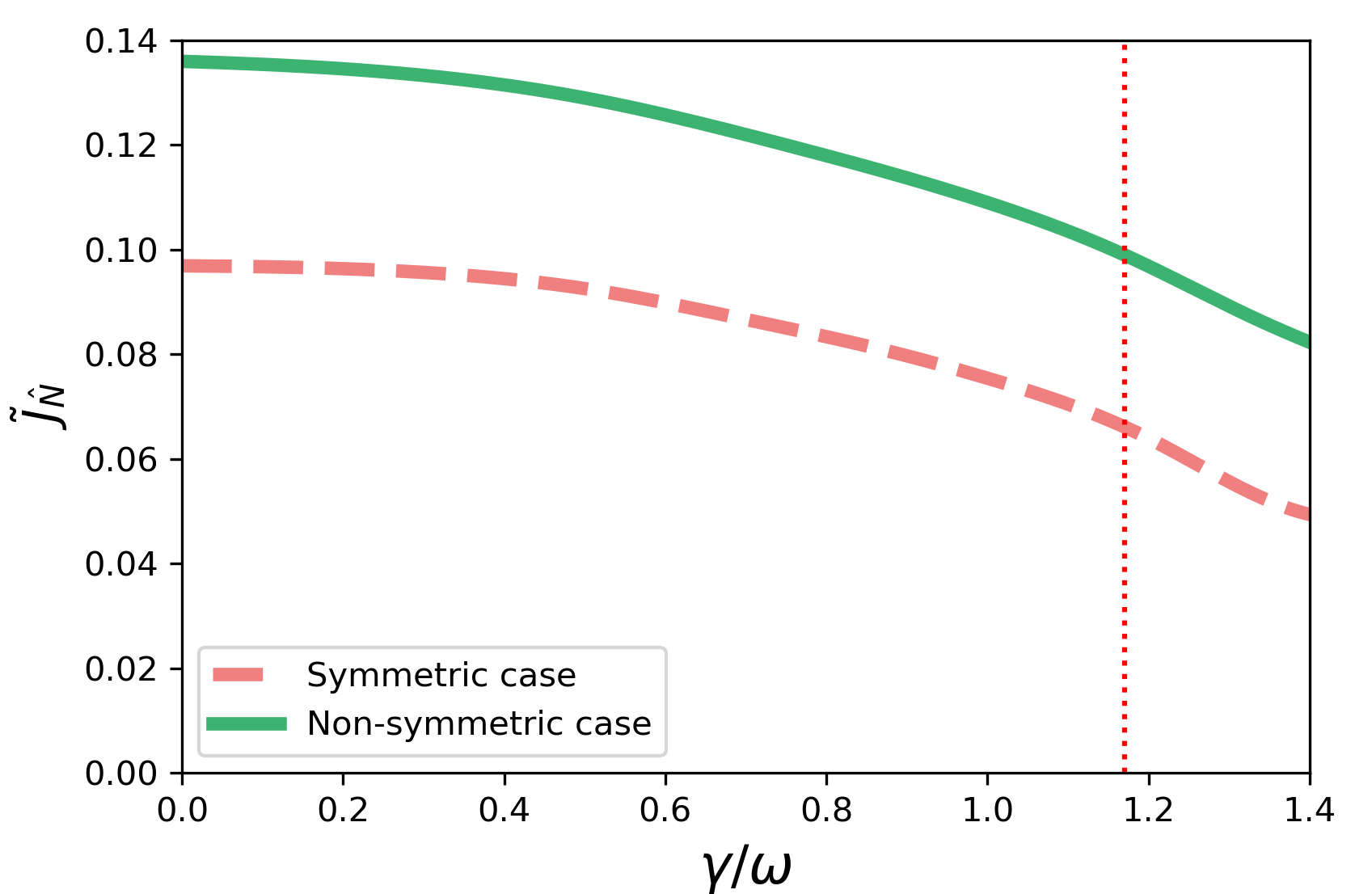}
\label{fig:fig14}
\caption{Current $\tilde{J}_{\hat{N}}$ as a function of $\gamma/\omega$ for $J=1$ and $\theta=0.52$. The~dashed red vertical line corresponds to $\gamma = 1.17$. The~system is decreasingly non-Markovian in the interval $0 < \gamma \lesssim 1.17$. For~$\gamma \gtrsim 1.17$, the~system is Markovian since we have $\gamma(t) > 0$. The~green line represents the non-symmetric case, while the dashed coral line represents the symmetric~one.}\label{asymetric2}
\end{figure}

The effect of increasing the positive contribution $\gamma$ is shown in Figure~\ref{asymetric2} for fixed $J$ and $\theta$ ($J=1$ and $\theta = 0.52$, for~which $\gamma(t)$ is plotted in Figure~\ref{Gammas}), where the solid green line is the plot of the current as a function $\gamma$ in the non-symmetric configuration.
The non-Markovian dynamics becomes Markovian for the value of $\gamma \approx 1.17$ correspondent to the dotted red line.
For $\gamma \gtrsim 1.17$, the~system is Markovian.
We see that the increase in the Markovian contribution, $\gamma$, cannot lead to an increase in the current, $\tilde{J}_{\hat{N}}$.
We note that the same effect -- the decrease of $\tilde{J}_{\hat{N}}$  as $\gamma$ increases -- is observed for other values of $J$ and $\theta$.
Therefore, the~decrease in the non-Markovianity of the system by increasing $\gamma$ jeopardizes the transport~efficiency.

\vspace{-12pt}

\subsection{Symmetric~Configuration}

We consider the case $\gamma = 0$ in Equation \eqref{gammat} in the symmetric configuration.
In \mbox{Figure~\ref{DAT}}, we have the current $\tilde{J}_{\hat{N}}$ plotted as a function of $\theta$, see the dashed coral line.
We see that non-Markovian dephasing-assisted transport does not happen, as~ we have a monotonic behavior of the current with $\theta$.
This behavior is similar to the cases discussed before for Figure~\ref{SIMSUM}, where we also have a monotonic behavior for the current in all the~cases.

Next, we study what happens when the positive contribution $\gamma$ is increased, by~taking $J=1$ and $\theta = 0.52$ once again, but~now in the symmetric configuration.
In Figure~\ref{asymetric2}, the~dashed coral line shows the current as function of $\gamma$.
As in the non-symmetric case, we see a monotonic decreasing behavior of the current $\tilde{J}_{\hat{N}}$ as $\gamma$ decreases.
Once again, the~non-Markovian dynamics becomes Markovian for $\gamma \approx 1.17$, indicated by the dotted red line in Figure~\ref{asymetric2}.
As in the non-symmetric case, therefore, non-Markovian scenarios are shown to be associated with a greater transport~efficiency.

\section{Conclusions}\label{conc}

The investigation of the influence of time-dependent dephasing on the efficiency of quantum transport is a very relevant problem in the context of open quantum systems research, notably in non-Markovian scenarios.
Here, we provided a systematic investigation of this phenomenon for a chain of coupled two-level systems, which is, in~turn, locally subjected to incoherent injection and extraction of energy in and out of the chain.
An exciton current is then established and evaluated when the system is in the stationary state. 

Specifically, the~models treated here are characterized by complete positive maps associated with time-dependent dephasing rates described by linear combinations of sine functions and constants in the canonical representation of the corresponding master equation.
Based on that, we studied the behavior of the exciton current in the non-symmetric and symmetric scenarios.
We found that the phenomenon of non-Markovian dephasing-assisted transport occurred in the non-symmetric cases, thereby, establishing a parallel with the time independent Markovian cases as investigated elsewhere~\cite{Elinor}.

Substantial advances for the understanding of the effect of Markovian and non-Markovian dephasing in the transport efficiency has been  made recently in the context of disordered chains~\cite{Novo, Elinor2, Zhang, Ancheyta, Hauke, Maier}. 
In \cite{Zhang}, for~instance, it was found that Markovian dephasing can help transport even for disordered symmetric chains. 
In \cite{Ancheyta}, it was found that the dephasing rate range in which dephasing-assisted transport occurs is significantly larger in the non-Markovian scenario than in the Markovian counterpart~\cite{Ancheyta}. 
Similarly, researchers \cite{Hauke}   found that non-Markovian dephasing can hold larger values for transport efficiency over a broader parameter range when compared to the Markovian case. 
These behaviors are qualitatively similar to the ones verified here, which indicates that non-Markovian dephasing may be a useful technique when it comes to transport~efficiency.

As a final remark, one can also investigate the so-called ``maximally non-Markovian evolution''  \cite{Budini}, to~find that different degrees of non-Markovianity in that model do not change the efficiency of quantum transport.
This indicates that the generalization of non-Markovian dephasing-assisted transport, beyond~the models studied here, is not straightforward, and~certainly deserves to be further investigated. We hope that our work will serve as a motivation for further advances related to this interesting~problem.
\\
\\
\textit{Acknowledgements} -- S.V.M. acknowledges support from the Brazilian agency CAPES.
S.V.M. acknowledges support from the Brazilian agency CAPES and the Knut and Alice Walenberg Foundation (KAW) (project 2016.0089). B.M. and F.L.S. acknowledge partial support from the Brazilian National Institute of Science and Technology of Quantum Information (CNPq-INCT-IQ 465469/2014-0) and CAPES/PrInt Process No. 88881.310346/2018-01. F.L.S. also acknowledges partial support from CNPq (Grant No. 305723/2020-0).


\end{document}